# EXACT SOLUTION OF TERZAGHI'S CONSOLIDATION EQUATION AND EXTENSION TO TWO/THREE-DIMENSIONAL CASES (3ᵀᴴ VERSION)

ROMOLO DI FRANCESCO

Wizard Technology, Teramo (TE) - Italy

web: www.romolodifrancesco.it - e-mail: romolo.difrancesco@vodafone.it

**ABSTRACT**

The differential equation by Terzaghi [1], Terzaghi and Fröhlich [2], better known as Terzaghi's one-dimensional consolidation equation, simulates the visco-elastic behaviour of soils depending on the loads applied as it happens, for example, when foundations are laid and start carrying the weight of the structure. Its application is traditionally based on Taylor's solution [3] that approximates experimental results by introducing non-dimensional variables that, however, contradict the actual behaviour of soils. After careful examination of the theoretical and experimental aspects connected with consolidation, the proposal of this research is a solution consisting in a non-linear equation that can be considered correct as it meets both mathematical and experimental requirements. The solution proposed is extended to include differential equations relating to two/three dimensional consolidation by adopting a transversally isotropic model more consistent with the inner structure of soils. Finally, this essay is complete with application examples that give more reliable results than the traditional solution. Future developments are also highlighted considering that the uniqueness theorem has not been proven yet.

## 1. INTRODUCTION TO THE HYDRAULIC AND MECHANICAL BEHAVIOUR OF SOILS

The mechanical behaviour of soils was coded only after the introduction of the "concept of effective stresses" [1] which marked the birth of Soil Mechanics starting from the general structure of Continuum Mechanics from which it derives. This concept is based on the inner structure of soils that are composed of a solid skeleton and inter-particle gaps. These pores are more or less interconnected and through them run fluids of different nature. Therefore, in view of a necessary simplification of the mathematics of associated phenomena, the concept of effective stresses requires the soils to be assimilated to bi-phasic systems composed of a solid skeleton saturated with water, *i.e.* two continuous means that act in parallel and share the stress status [4]:

$$\sigma'_{ij} = \sigma_{ij} - u_0 \delta_{ij} \qquad (1)$$

In equation (1) there is the tensor of the total stresses exerted by the solid skeleton ($\sigma_{ij}$), the hydrostatic pressure exerted by the fluid ($u_0$, known as interstitial pressure) and Kronecker's delta ($\delta_{ij}$); furthermore, from a merely phenomenological point of view, equation (1) attributes the soil shear resistance only to effective stress, independent of the presence of the fluid.

It should be highlighted that the concept of effective stresses is valid only in stable conditions, when the fluid is in balance with the solid skeleton. In these conditions you can calculate the hydrostatic component in all points of the underground and in all moments through the application of the laws of balance; vice-versa, in transient conditions, it is necessary to introduce other elements capable of accounting for the variation of the component $u$ induced by stresses of various nature.

At this point, the problem focus is on the permeability coefficient ($K = m/sec$) that, by expressing the capacity of a soil to transmit a fluid, takes on the character of a velocity and varies approximately in the range $10^{-1} \div 10^{-10}$ m/sec depending on the inner structure of the solid skeleton. As a direct consequence of this extreme variability, the soils with high permeability (such as sands) behave as open hydraulic systems where compression induced, for example, by the load of a foundation, causes simultaneous drainage of the fluid from the pores. In practice, the fluid does not take part in the mechanical response and the stress induced weighs only on the solid skeleton that, in turn, subsides in association with reduced porosity. On the contrary, soils with very low permeability (such as clays) exhibit hydraulic delay in reacting to stresses, with consequent development of an initial interstitial overpressure ($u \neq 0$) that contradicts equation (1), and participate in the mechanical response with the solid skeleton. A transient filtering motion follows that comes to end only when the initial value of the interstitial pressure is reset ($u = 0$).

In practice, given that a deformation of the solid skeleton occurs together with the expulsion of water, sands develop elasto-plastic settlements synchronous with load application while clays exhibit time-dependent consolidation settlements typically characterized as reverse hyperbolic functions.

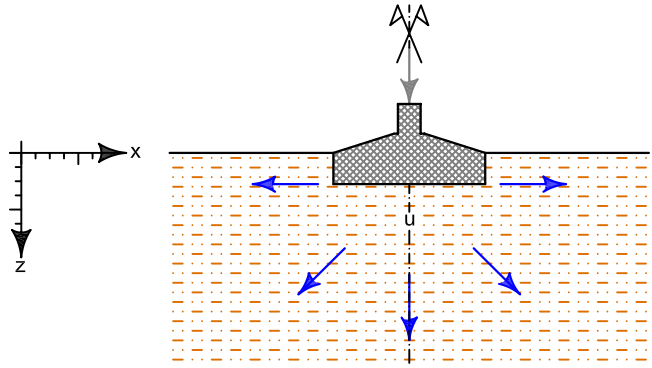

Figure 1. The load transmitted by a foundation always causes interstitial overpressures whose dissipation depends on soil permeability that in turn depends on porosity. Consolidation may last from some minutes (loose sands) to tens of years (very stiff clays).

## 2. INTRODUCTION TO THE TERZAGHI'S THEORY OF ONE-DIMENSIONAL CONSOLIDATION

The one-dimensional consolidation equation [1, 2] describes the hydraulic behaviour of soils in transient conditions by making it possible to simulate the variation in time of interstitial overpressures ($u$), generated – for example – by the load induced by a foundation or by a road embankment (figure 1), with consequent visco-elastic settlements to which corresponds a structural reorganisation of the solid skeleton, with reduction of porosity and, concurrently, of the degrees of freedom.

This formulation can be inferred from applying the continuity equation to supposedly saturated soils leading [5], with a few mathematical manipulations, to the following relationship that demonstrates how the transient filtering motion depends on the vertical permeability factor ($K_z$), on the compressibility factor ($m_v$) and on the weight of the volume of water ($\gamma_w = 10\ kN/m^3$) that in turn identifies the fluid:

$$\frac{1}{m_v \cdot \gamma_w} \cdot \left( K_z \cdot \frac{\partial^2 u}{\partial z^2} + \frac{\partial K_z}{\partial z} \cdot \frac{\partial u}{\partial z} \right) = \frac{\partial u}{\partial t} \tag{2}$$

The next step consists in introducing the hypothesis (in contrast with experimental results) that the permeability factor does not change during consolidation:

$$\frac{1}{m_v \cdot \gamma_w} \cdot \left( K_z \cdot \frac{\partial^2 u}{\partial z^2} \right) = \frac{\partial u}{\partial t} \tag{3}$$

Finally, denoting the consolidation coefficient as $c_v$:

$$c_v = \frac{K_z}{m_v \cdot \gamma_w} = \frac{K_z \cdot E_{ed}}{\gamma_w} \tag{4}$$

you come to write the classical one-dimensional consolidation equation:

$$c_v \frac{\partial^2 u}{\partial z^2} = \frac{\partial u}{\partial t} \tag{5}$$

It should be noticed that equation (5) is analogous to Fourier's law on heat propagation to the point that you can define the theory of consolidation as the simulation of the propagation of stress-induced interstitial pressures in the subsoil.

### 2.1 TAYLOR'S SOLUTION TO THE ONE-DIMENSIONAL CONSOLIDATION THEORY

The solution of equation (5) may be found introducing two non-dimensional variables:

$$Z = \frac{z}{H}$$
$$T_v = \frac{c_v \cdot t}{H^2} \tag{6}$$

Both expressed as a function of $H$ that is the maximum drainage path of the fluid.
In this manner, equation (5) becomes:

$$\frac{\partial^2 u}{\partial Z^2} = \frac{\partial u}{\partial T_v} \qquad (7)$$

whose analytical solution is [3]:

$$u(z,t) = \sum_{m=0}^{\infty} \frac{2u_{(0)}}{M} (\sin MZ) \cdot e^{-MT_v} \qquad (8)$$

$$M = \frac{\pi}{2}(2m+1) \qquad (9)$$

valid for $m \in N$ e based on the assumption that $u$ remains constant with depth.

According to this solution, the application of a load causes an interstitial overpressure in the soil (initial conditions: t = 0, $u_{(0)}$ = $u_0$ with 0 ≤ Z ≤ 2), that rapidly drops to zero at the drainage surfaces (contour conditions: t>0, u(Z=0) = u(Z=2) = 0), generating an isochrone (regarding the concept of effective stresses in a transient phase) whose bottom coincides with the central line of the consolidating layer that, in turn, can be considered as a waterproof surface (figure 2a).

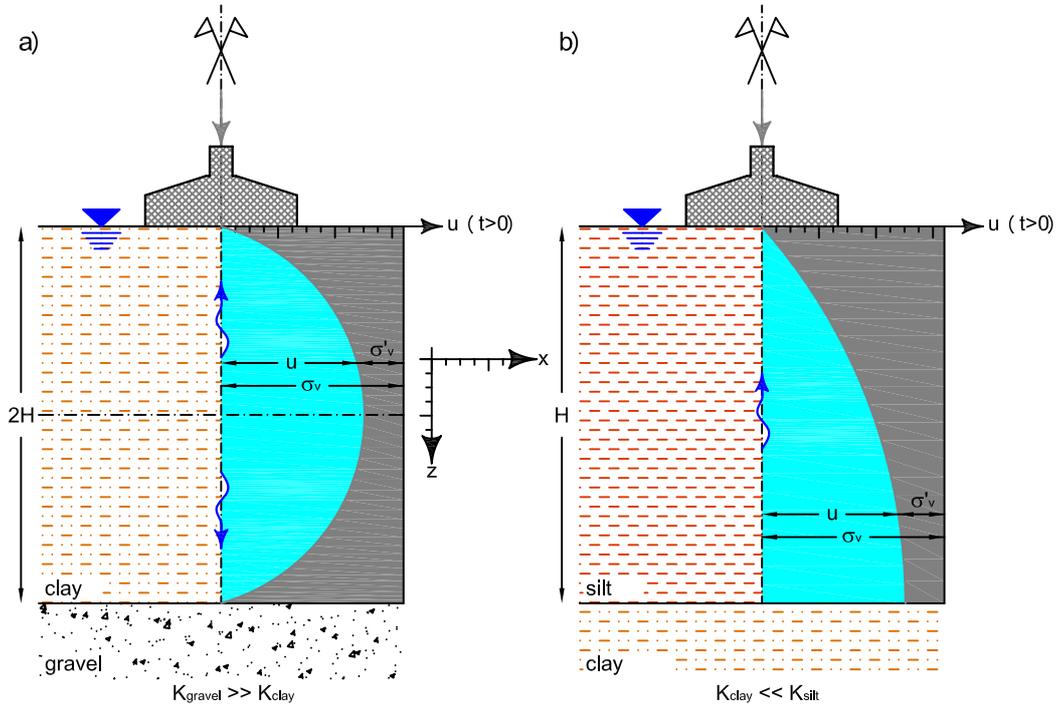

Figure 2. Definition of the draining thickness and of its interstitial overpressure isochrone as a function of relative permeability between contacting layers.

On the basis of the above reasoning, the solution can be extended also to the cases with drainage at one edge, provided that attention is paid at identifying the correct value of $H$ (figure 2b).

Since drainage resulting from consolidation leads to yielding, it is also possible to define the average consolidation degree:

$$U_m = \frac{s(t)}{s_f} \qquad (10)$$

expressed as a function of yielding in an instant ($s(t)$) relative to final yielded ($s_f$). Furthermore, also the following relationship is demonstrated to be valid:

$$U_m = 1 - \sum_{m=0}^{\infty} \frac{2}{M^2} \cdot e^{-M^2 T_v} \qquad (11)$$

whose solution is shown in figure 3 as a function of $T_v$.

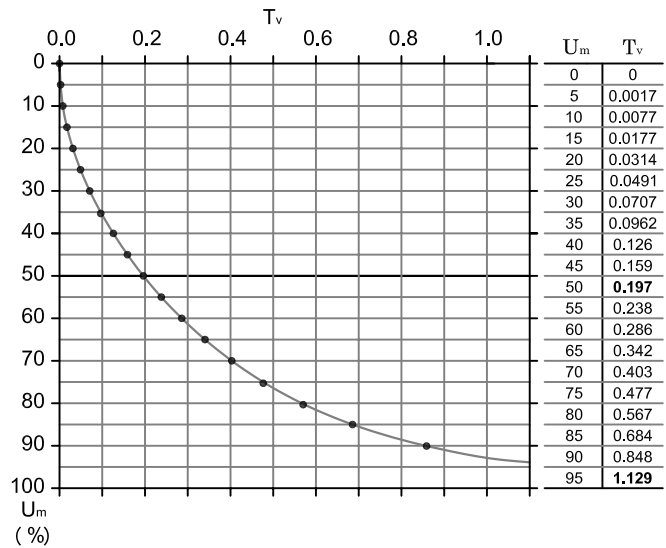

Figure 3. $T_v$ variation curve as a function of the consolidation degree $U_m$ and associated table of the values coded by Taylor's solution (source: [6] – with modifications).

Alternatively, you can use the approximate solutions by Sivaram and Swamee [7]:

$$U_m = \frac{(4T_v/\pi)^{0,5}}{\left[1+(4T_v/\pi)^{2,8}\right]^{0,179}}$$

$$T_v = \frac{(\pi/4)\cdot(U_m)^2}{\left[1-(U_m)^{5,6}\right]^{0,357}}$$

(12)

To conclude, Taylor's solution corresponds to a development in series of Fourier truncated at the top, which describes consolidation through a reverse hyperbolic function coded through fixed values of $T_v$ expressed as a function of $U_m$.

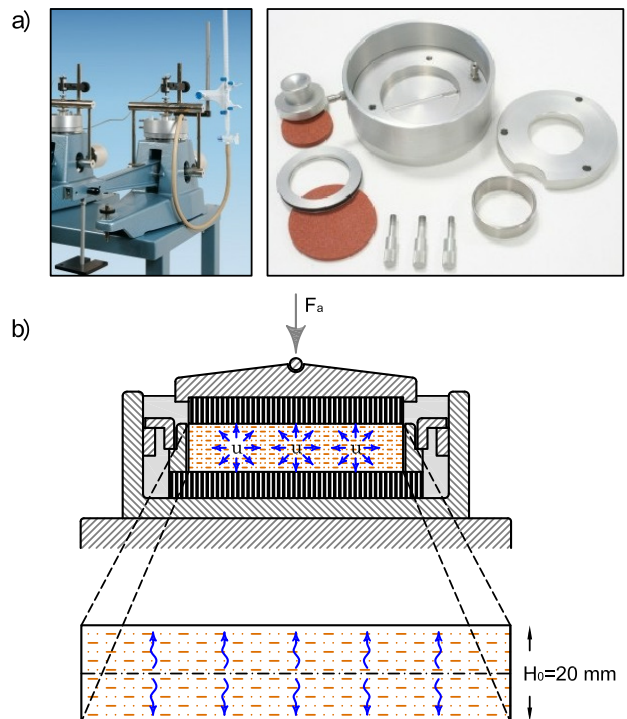

Figure 4. Details of an oedometric cell (a); section with magnification of the test-piece and indication of the drainage paths (b).

## 2.2 Experimental definition of the theory of consolidation

The application of equation (5) is linked to the execution of oedometric tests (figure 4), whose first prototype dates back to Terzaghi [8], Casagrande [9], Gilboy [10] and Rutledge [11]. Only vertical stresses and deformations ($\sigma'_z$, $\varepsilon_z$) can be assessed in these tests because side deformations ($\varepsilon_r$) are actually impossible and radial stresses ($\sigma'_r$) cannot be measured. In this manner, its constitutive equation can be written and expressed in differential terms, in the following simplified form:

$$\left\{\begin{array}{c} d\varepsilon_z \\ 0 \end{array}\right\} = \left[\begin{array}{cc} C_{aa} & C_{ar} \\ C_{ra} & C_{rr} \end{array}\right] \cdot \left\{\begin{array}{c} d\sigma'_z \\ d\sigma'_r \end{array}\right\} \qquad (13)$$

where appears the compliance matrix $[C]$, valid for oedometric conditions. See details in [12, 13].
Equation (13) can also be written in the extended form:

$$d\varepsilon_z = C_{aa} \cdot d\sigma'_z + C_{ar} \cdot d\sigma'_r \qquad (14a)$$

$$0 = C_{ra} \cdot d\sigma'_z + C_{rr} \cdot d\sigma'_r \qquad (14b)$$

Therefore, the $d\sigma'_r$ expression can be inferred from equation (14b), as follows:

$$d\sigma'_r = -\frac{C_{ra}}{C_{rr}} \cdot d\sigma'_z \qquad (15)$$

that in turn can be introduced into equation (14a):

$$d\varepsilon_z = d\sigma'_z \cdot \left(C_{aa} - \frac{C_{ar} \cdot C_{ra}}{C_{rr}}\right) \qquad (16)$$

If you analyse equations (15) and (16), you will find out that the former provides the formal relationship of the vertical and radial stresses while the latter demonstrates that vertical deformations depend on the combination of axial and radial visco-elastic constants.

In brief, as far as the phenomenological meaning of the above formulations is concerned, oedometric tests can be fruitfully used to analyse the one-dimensional effects induced by an unspecific stress assuming that the conditions of geometrical symmetry illustrated in figures 1 and 3 are valid, and taking into account that equation (16) can be also written as follows:

$$\frac{d\varepsilon_z}{d\sigma'_z} = \left(C_{aa} - \frac{C_{ar} \cdot C_{ra}}{C_{rr}}\right) = \frac{1}{E_{ed}} = m_v \qquad (17)$$

In this manner, you work out the formulation of the elastic module in oedometric conditions ($E_{ed}$) that, in turn, is in inverse proportion to the compressibility factor ($m_v$) shown in equation (4).

Finally, it should be noticed that the oedometric condition is a special case of flat deformation, in the absence of side deformations and corresponds to a null value of Poisson's factor ($\nu = 0$), associated only to the existence of geometrical symmetries of the type shown in figure 1.

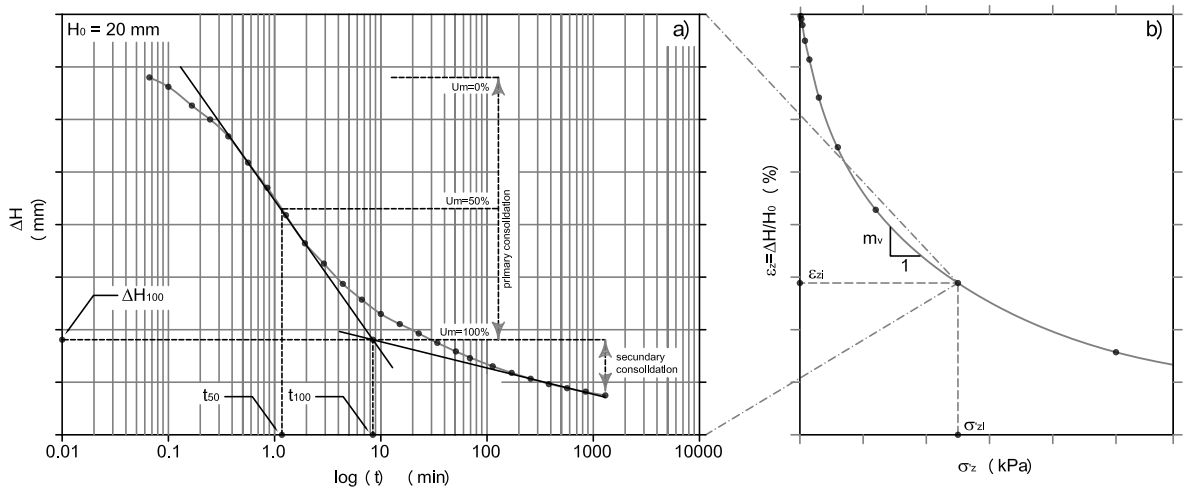

Figure 5. Example of a consolidation curve for a certain load condition $\sigma'_{z,i}$ (a) and interpretation of final results in the plane $\varepsilon_z \div \sigma'_z$ (b).

## 2.3 EXPERIMENTAL DETERMINATION OF THE CONSOLIDATION COEFFICIENT

The application of the theory of consolidation is strictly connected to the determination of $c_v$, coded according to two methods, both connected with the oedometric tests [14]. These tests consist in subjecting a round test piece [15], confined into an indeformable ring (figure 4), to axial loads according to a loading schedule ($F_a$) following a geometric progression $q$ like:

$$F_{a(n)} = q \cdot F_{a(n-1)} \qquad (18)$$

Each load is held for a time that guarantees the passage from load-dependent primary consolidation to secondary consolidation that depends on the merely rheological behaviour of soils.

The most reputed methodology [16] requires the identification of the following elements on the experimental curves (figure 5a):

1) $t_{100}$: time necessary for conventional completion of primary consolidation;
2) $t_{50}$: time to attain the condition $U_m = 50\%$, equation (10), in relation with instrumental double drainage of the test-piece (figure 4);
3) $H_{100}$: height of the test-piece at the end of conventional primary consolidation, calculated starting from initial height $H_0 = 20$ millimeters from which the value $\Delta H_{100}$ read in the diagram is subtracted.

In this manner, the second equation (6) can be reversed:

$$c_v = \frac{T_v \cdot (H_{100}/2)^2}{t_{50}} = \frac{0,197 \cdot H_{100}^2}{4 t_{50}} \qquad (19)$$

as a function of both $T_{v(50\%)} = 0.197$ (figure 3) and semi-height $H_{100}/2$ of the test piece in each loading step applied (figure 5b).

Known $c_v$ for the design load (meaning that it should be selected depending on the stress level expected), the next step is the calculation of the time of completion of consolidation:

$$t = \frac{T_{v(95\%)} \cdot H^2}{c_v} = \frac{1,129 \cdot H^2}{c_v} \qquad (20)$$

being $T_{v(95\%)} = 1.129$ (figure 4) and $H$ the height of the layer according to that highlighted in figure 2.

Finally, known the consolidation settlements (that can be calculated using Geotechnics methods based on the interpretation of the oedometric tests), you can build the time-settlement curve for the different structure/foundation nodes from which to obtain the differential settlement variation field (figure 6).

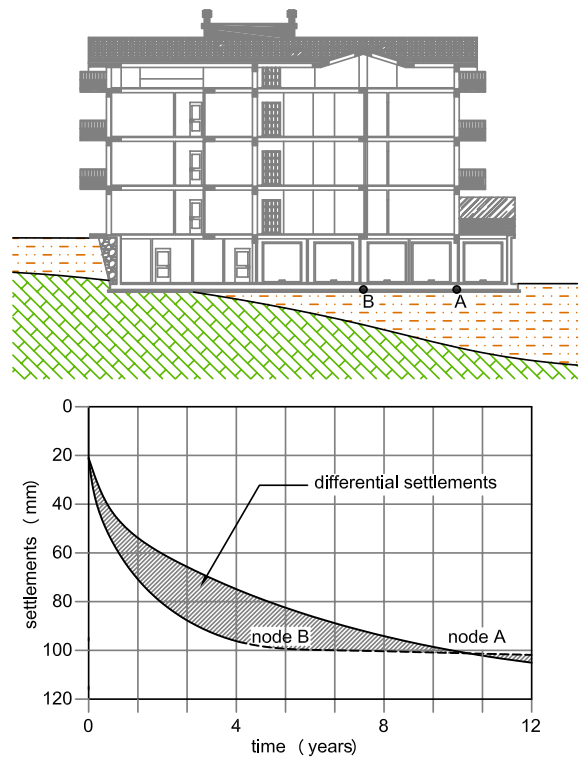

Figure 6. Example of absolute settlement variation in time relating to two nodes of the same foundation (source: [17] – with modifications).

## 2.4 LIMITS OF TAYLOR'S SOLUTION

If you analyse Taylor's solution [3], you will find out that its application relies on using the non-dimensional variable $T_v$, whose values are coded independent of the soil analysed, and that is used both in experimental determination of the consolidation coefficient ($c_v$) and in the calculation of the consolidation completion time. At this point, a problem arises from experimental observation, namely the mechanical behaviour of a soil depending on its geological history (that is, a tenso-deformative history), it cannot be referred to the fixed values of a non-dimensional variable. On the other hand, it suffices to reflect on clay formation environments, which depend on hydrolysis which in turn is influenced by temperature and water, *i.e.* by latitude, altitude and the climate system, to understand the intrinsic limits of using the parameter $T_v$.

Also the construction of the time-settlement curve is quite difficult as it requires the following passages:

- calculation of time ($t_i$) for each value of $U_m$, by introducing the corresponding values of $T_v$ from figure 3 into equation (20);
- product of the final settlement ($s_f$) and the single percentage values of $U_m$ from which to obtain the corresponding values of settlement $s_i$;
- assembly of the diagram $s_i$-$t_i$ (figure 6).

In the end, being not possible to simulate the real mechanical and hydraulic behaviour of soils correctly, we consider the Taylor's solution as experimentally inadequate. In other words, the exact solution of equation (5) would depend exclusively on the consolidation coefficient $c_v$ that, expressed as a function of $K_z$ and of $m_v$, is a real constitutive link.

To conclude, it should also be considered that, once determined $c_v$ for each load step applied, the experimental practice foresees the calculation of the corresponding $K_z$, taken from equation (4), being known the value of $m_v$ which is obtained by applying equation (17) to the experimental curves (figure 5a). However, using this procedure includes that, due to the remarkable non-linearity of the compressibility factor, $K_z$ markedly depends on the stress level and, consequently, on the deformative level. Furthermore, since the method is connected with the consolidation velocity through a direct relationship with $c_v$, rather inaccurate measures are obtained due to the physical magnitudes that play a role in the process as well as to their non-linearity.

## 3. EXACT SOLUTION OF TERZAGHI'S CONSOLIDATION EQUATION

Let's assume that $u, c_v, k_z > 0$ are three positive constants assigned and that:

$$u : [0, +\infty) \times [0, +\infty) \to \Re \tag{21}$$

is the regular function given by:

$$u(z,t) = u \cdot e^{-k_z z} \cos\left(2 c_v k_z^2 t - k_z z\right) \tag{22}$$

Now, you can easily notice that $u(z,t)$ solves the differential equation (5); indeed, given the validity of the following:

$$\frac{\partial u}{\partial t} = -2 c_v k_z^2 u e^{-k_z z} \sin\left(2 c_v k_z^2 t - k_z z\right) \tag{23}$$

$$\frac{\partial u}{\partial z} = -k_z u e^{-k_z z} \cos\left(2 c_v k_z^2 t - k_z z\right) + k_z u e^{-k_z z} \sin\left(2 c_v k_z^2 t - k_z z\right) \tag{24}$$

$$\frac{\partial^2 u}{\partial z^2} = -2 k_z^2 u e^{-k_z z} \sin\left(2 c_v k_z^2 t - k_z z\right) \tag{25}$$

you obtain:

$$c_v \frac{\partial^2 u}{\partial z^2} = -2 c_v k_z^2 u e^{-k_z z} \sin\left(2 c_v k_z^2 t - k_z z\right) = \frac{\partial u}{\partial t} \tag{26}$$

which is the differential equation given.

If you analyse equation (22), you will find out that it simulates the time variation of interstitial overpressures in the subsoil through the consolidation constant $k_z$ – starting from the point where they are triggered – by dampening their width as depth increases through a reverse hyperbolic function.

Now, since equation (22) is the solution of equation (5), it should be necessarily extended also to experimental methods – oedometrically considered – in order to determine the parameters that govern it correctly. In this sense, it may be useful to analyse an important property of the function $u$.

Given $H > 0$ (figure 4), it can be useful to identify the instant $t_H > 0$ to which the following conditions apply:

$$u(H, t_H) = 0, \quad u(H, t) \neq 0, \quad \text{if} \quad 0 < t < t_H \tag{27}$$

As a first passage, you should notice that equation $u(H,t) = 0$ reduces to the form:

$$\cos(2c_v k_z^2 t - k_z H) = 0 \qquad (28)$$

which has infinite solutions like:

$$2c_v k_z^2 t - k_z H = \frac{\pi}{2} + h\pi \quad \text{upon variation of } h \in Z \qquad (29)$$

or like:

$$t = \frac{2k_z H + \pi}{4c_v k_z^2} + \frac{\pi}{4c_v k_z^2} h \quad \text{upon variations of } h \in Z \qquad (30)$$

The next passage consists in selecting the positive value $t$ closest to zero from among those determined by setting the condition $t > 0$ that provides:

$$\frac{2k_z H + \pi}{4c_v k_z^2} + \frac{\pi}{4c_v k_z^2} h > 0 \qquad (31)$$

from which we obtain:

$$h > -\frac{2k_z H + \pi}{2\pi} \qquad (32)$$

The last passage includes that, if you set also the following condition:

$$h_H = \min\left\{ h \in Z : h > -\frac{2k_z H + \pi}{2\pi} \right\} \qquad (33)$$

you obtain the formulation of the consolidation completion time:

$$t_H = \frac{2k_z H + \pi + 2h_H \pi}{4c_v k_z^2} \qquad (34)$$

Finally, from equation (34) you can extract the consolidation coefficient:

$$c_v = \frac{2k_z H + \pi + 2h_H \pi}{4t_H k_z^2} \qquad (35)$$

Now it should be noticed that for equation (35) to allow the determination of $c_v$ starting from experimental data, the additional parameter $k_z$ introduced in the initial hypotheses must be defined correctly. To this regard, considering that equation (5) describes the propagation of interstitial overpressures in the subsoil, it follows that the solution proposed – equation (22) – can also be written in a fully equivalent form that satisfies both the definition given and the passages corresponding to equations (23) to (26):

$$u(z,t) = u \cdot e^{-k_z z} \cos(\omega t - k_z z) \qquad (36)$$

after setting:

$$2c_v k_z^2 = \omega \qquad (37)$$

as a function of the angular frequency $\omega = (2\pi)/t$.
The parameter sought can be taken from equation (37):

$$k_z = \sqrt{\frac{\omega}{2c_v}} = \sqrt{\frac{\pi}{c_v t}} \qquad (38)$$

In accordance with the work hypotheses, it depends exclusively on the consolidation coefficient as it should not be determined or added as an external condition to differential equation (5).
Now equation (38) can be introduced into equation (35) and with a few mathematical manipulations you obtain:

$$c_v = \frac{4H^2}{\pi t_H (15 - 4h_H^2)} \qquad (39)$$

What is left to do now is just to assess correspondence of equation (39) and the experimental data that can be obtained from oedometric tests by setting the following conditions as a function of figure 4 and of the diagram in figure 5a:

$$H = H_0, \quad h_H = \frac{H_{100}}{2}, \quad t_H = t_{50} \qquad (40)$$

These conditions, once introduced into equation (39), give the parameter sought without adding any further hypotheses or additional variables:

$$c_v = \frac{4H_0^2}{\pi t_{50}\left(15 - 2H_{100}^2\right)} \tag{41}$$

As an example, with $H_0$ = 0.02 meters, $t_{50}$ = 912 seconds and $H_{100}$ = 0.01668 meters, obtained from a real-life test, you obtain:
- equation (19) → $c_v$ = 1.5×10⁻⁸ m²/s
- equation (41) → $c_v$ = 3.7×10⁻⁸ m²/s

To conclude, the solution proposed with equation (22) can be considered correct even though the uniqueness theorem has not been proven yet, since it perfectly corresponds both mathematically and experimentally. Obviously, by analogy, the same solution can be extended also to Fourier's equation on heat transmission, after replacing the consolidation coefficient with the thermal diffusivity coefficient ($c_v \equiv \alpha$) and temperature ($u \equiv T$) to interstitial overpressures.

## 3.1 EXTENSION TO THE TWO-DIMENSIONAL AND THREE-DIMENSIONAL CASES

Let's put $u, c_x, c_z, k_x, k_z > 0$ to be positive constants assigned and be:

$$u : [0, +\infty)^2 \times [0, +\infty) \to \Re \tag{42}$$

the regular function given by:

$$u(x,z,t) = u \cdot e^{-k_x x - k_z z} \cos\left[2\left(c_x k_x^2 + c_z k_z^2\right)t - k_x x - k_z z\right] \tag{43}$$

Going through the same passages as in equations (23) to (26), you can easily notice that $u(x,z,t)$ solves the differential equation:

$$c_x \frac{\partial^2 u}{\partial x^2} + c_z \frac{\partial^2 u}{\partial z^2} = \frac{\partial u}{\partial t} \tag{44}$$

Similarly, if $u, c_x, c_y, c_z, k_x, k_y, k_z > 0$ are positive constants assigned and:

$$u : [0, +\infty)^3 \times [0, +\infty) \to \Re \tag{45}$$

the regular function given, then:

$$u(x,z,t) = u \cdot e^{-k_x x - k_y y - k_z z} \cos\left[2\left(c_x k_x^2 + c_y k_y^2 + c_z k_z^2\right)t - k_x x - k_y y - k_z z\right] \tag{46}$$

solves the differential equation:

$$c_x \frac{\partial^2 u}{\partial x^2} + c_y \frac{\partial^2 u}{\partial y^2} + c_z \frac{\partial^2 u}{\partial z^2} = \frac{\partial u}{\partial t} \tag{47}$$

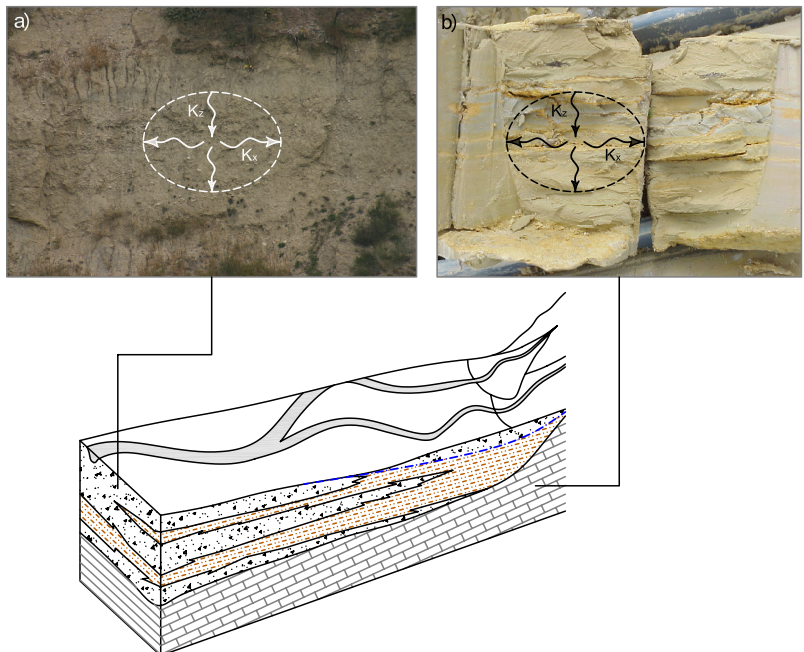

Figure 7. Examples of hydraulic anisotropy as a function of the soil inner structure: (a) fluvial deposits; (b) sedimentary basins.

Considering the soils' markedly anisotropic hydraulic and mechanical behaviour that can be attributed to the structure of the solid skeleton that in turn depends on accumulation characteristics (figure 7), the application of the solution to the 2D case - equation (43) – does not necessarily requires the preliminary determination of all parameters introduced. Mathematically speaking, the geological development of sedimentary beds leads to assimilating soils to transversally isotropic means to which the following conditions apply:

$$K_x > K_z \tag{48}$$

$$\frac{K_x}{K_z} = m \quad \rightarrow \quad K_x = mK_z \tag{49}$$

Using equation (49) requires the experimental determination of permeability in the horizontal plane ($K_x$), such as in the three-axial cells or permeameters, by applying the rotation of the axes to the test-pieces. It follows that the relationship in equation (4) can be extended also to the consolidation coefficients:

$$\frac{c_x}{c_z} = m \quad \rightarrow \quad c_x = mc_z \tag{50}$$

and to consolidation constants:

$$\frac{k_x}{k_z} = m \quad \rightarrow \quad k_x = mk_z \tag{51}$$

Finally, by repeating the same procedure as in equations (36) and (37), you obtain:

$$c_x k_x^2 + c_z k_z^2 = \frac{\omega}{2} \tag{52}$$

where to introduce equations (50) and (51) that lead initially to:

$$k_z = \sqrt{\frac{\pi}{c_z(m^3+1)t}} \tag{53}$$

and then, by introducing equation (53) into equation (51), to the determination of the parameter sought:

$$k_x = m\sqrt{\frac{\pi}{c_z(m^3+1)t}} \tag{54}$$

Notice that with the condition $m = 1$, equation (54) reduces to:

$$k_x = \sqrt{\frac{\pi}{2c_z t}} \quad \rightarrow \quad k_x = \sqrt{\frac{1}{2}}k_z \tag{55}$$

The procedure outlined can be fruitfully extended even to the 3D case by writing in sequence:

$$K_x = mK_z, \quad K_y = nK_z \tag{56}$$

$$c_x = mc_z, \quad c_y = nc_z \tag{57}$$

$$k_x = mk_z, \quad k_y = nk_z \tag{58}$$

$$c_x k_x^2 + c_y k_y^2 + c_z k_z^2 = \frac{\omega}{2} \tag{59}$$

$$k_z = \sqrt{\frac{\pi}{c_z(m^3+n^3+1)t}} \tag{60}$$

$$k_x = m\sqrt{\frac{\pi}{c_z(m^3+n^3+1)t}} \tag{61}$$

$$k_y = n\sqrt{\frac{\pi}{c_z(m^3+n^3+1)t}} \tag{62}$$

Furthermore, since the horizontal plane symmetry properties of transversal isotropic means $K_x = K_y$ and $k_x = k_y$ apply, equations (60), (61) and (62) can be simplified:

$$k_z = \sqrt{\frac{\pi}{c_z(2m^3+1)t}} \tag{63}$$

$$k_x = m\sqrt{\frac{\pi}{c_z(2m^3+1)t}} \quad (64)$$

$$k_y = n\sqrt{\frac{\pi}{c_z(2m^3+1)t}} \quad (65)$$

To conclude, with the condition *m = n = 1*, the equations above reduce to:

$$k_x = k_y = k_z = \sqrt{\frac{\pi}{3c_z t}} \quad (66)$$

i.e. to:

$$k_x = k_y = \sqrt{\frac{1}{3}}k_z \quad (67)$$

## 3.2 PRACTICAL APPLICATIONS

At the time being, the applications of the solutions provided by equations (22), (43) and (46) led to the development of two methods to analyse consolidation development over the time.

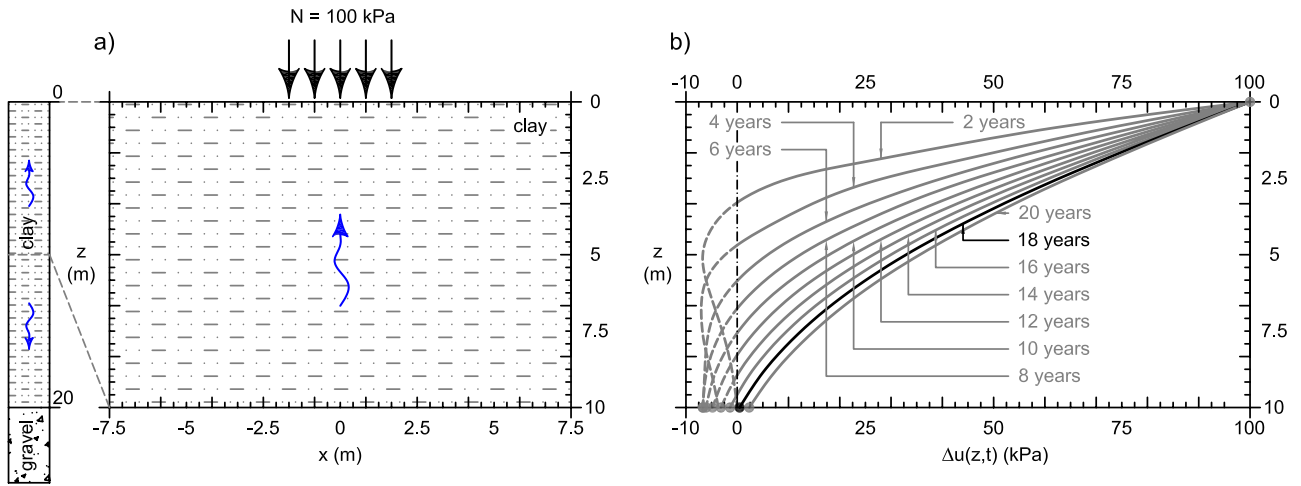

Figure 8. Application example of the solution of Terzaghi's differential equation: (a) reference stratigraphy; (b) dissipation curves.

As a first example, reference is made to the one-dimensional case given in equation (22). Consider a bank of clay, with 2H = 20 meters, resting on gravels and making up a drainage path with H = 10 meters (figure 8a); on this bank a static load N = 100 kPa is applied. The oedometric tests carried out (see figure 5a) gave $H_0$ = 0.02 meters (20 mm), $t_{50}$ = 145 seconds and $H_{100}$ = 0.019365 meters (19.365 mm) from which the following was obtained:
- $c_v$ = 0.011 m²/day – equation (19); $t_{95}$ = 28.1 years – equation (20);
- $c_v$ = 0.020 m²/day – equation (41).

Furthermore, let's assume that the action of the load results in an interstitial overpressure u = N = 100 kPa, relating with the extremely low permeability of clay, which involves that the stresses are entirely transferred to the fluid (concurrently with load application).

The first application of equation (22) consists in building *n* dissipation curves of *u* (figure 8b) in the range $t_1$ = 2 years to $t_n$ = 20 years. Now, if you analyse the diagram in detail, you will find out that consolidation completes in times around the 18-year curve ($U_m$ = 100%) because it respects the conditions given in equation (27). In this manner, even if you are going to analyse the one-dimensional field only, the settlement overestimation problem intrinsic to Taylor's solution for which applies $t_{95}$ = 28.1 years, is solved.

The second application consists in searching the time value that voids equation (22):

$$u(H, t_{100}) = u \cdot e^{-k_H H} \cos(2c_v k_H^2 t_{100} - k_H H) = 0 \quad (68)$$

If applied to the example in figure 8, this method gives the exact time of consolidation completion *$t_{100}$ = 17.5 years*.

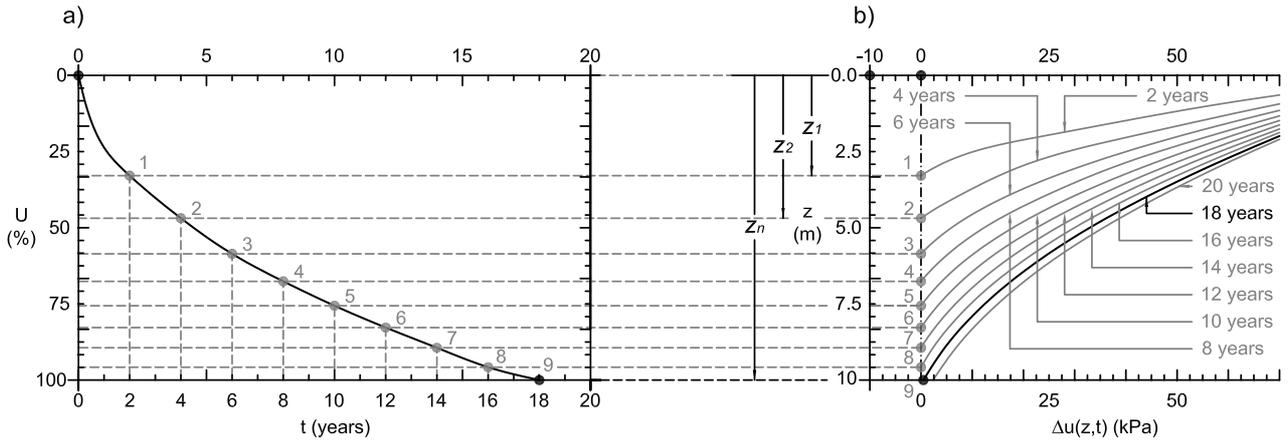

Figure 9. (a) comparison of the $U_m \div t$ curves for both the solution proposed and Taylor's solution; (b) identification of the $u$-voiding depths for each dissipation curve.

Then, the calculation is achieved by building the curve $U_m \div t$, noticing that each dissipation curve has a matching depth $z_i$ that voids the corresponding $u$ (figure 9b), as it happens with the consolidation completion curve $t_{z=10}$ = 17.5 years. Therefore, considering the following should apply:

$$U_{m,i(\%)} = \frac{z_i}{z_n} \cdot 100 \qquad (69)$$

finally you can determine $U_m$ values for each dissipation curve, which leads to the diagram in figure 9a.

However, if the application of the one-dimensional solution is immediate, with 2D consolidation it is necessary to identify the relationships expressed by equations (48) to (51) beforehand. Consequently, in order to evaluate how those parameters influence interstitial overpressure dissipation times, analyses were conducted presuming:

- m = 1, 2, 5, 10
- n = 0,1H, 0,5H, H, 2H,

as the horizontal dimension was not known *a priori*.

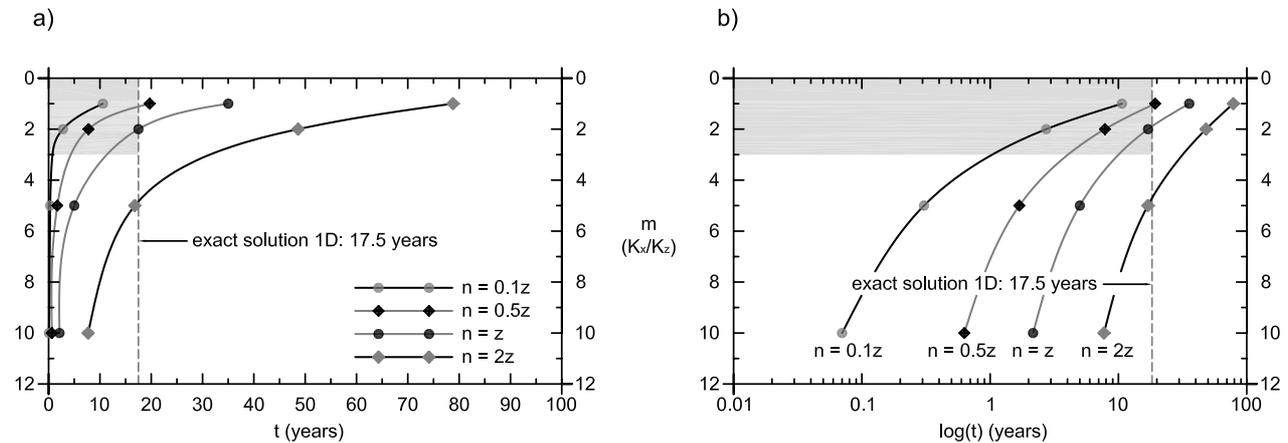

Figure 10. 2D solution of the example in figure 8a upon the variation of *m* and *n* represented in: (a) linear scale; (b) semi-logarithmic scale.

The results (figure 10a) demonstrate that as *m* increases, consolidation times decrease for each value of *n* taken, in line with the behaviour expected for a two-dimensional means. At the same time, as *n* increases, you observe that consolidation completion times increase and this can be attributed to a longer drainage path.

If you work out a diagram with the results in the $m \div log(t)$ plane (figure 10b), you perfectly realise the influence exerted by both *m* and *n*, considering the solution of the one-dimensional problem as a upper limit. This condition, matched with the experimental value of *m*, allows defining the range of action of the most probable value of *n*: see the grey area in figure 10 for a hypothetical m = 3. In other words, for a horizontal layer infinitely extended in that direction, the value of *n* to be introduced for the correct solution, depend on foundation size and consequent stress level induced in the subsoil.

## 4. CONCLUSIONS

With in mind the entire Mechanics of Soils, and the study of the soil visco-elastic behaviour [1, 2] in particular, the application of Terzaghi's differential equation, is historically based on Taylor's solution [3] that approximates experimental results – limited to the one-dimensional case only – through the introduction of arbitrary and fixed non-dimensional variables, independent of the geological history of the means.

After accurate examination of the theory and of the historical solution, and interpretation of lab data, this research work makes a proposal for a solution that can be considered correct as it solves the differential equation and, at the same time, allow correct interpretation of experimental data. Then, solution has been fruitfully extended to the two- and three-dimensional cases and finally tested in a real-life case.

To conclude, results even more satisfactory have come to light from the analysis of data. At the same time, they have opened additional research channels, considering that the uniqueness theorem has not been proved yet and that the influence of underground geometry should evaluated in the two- and three-dimensional cases. Monitoring of future geotechnical structures and/or retrospective analysis of real-life cases may clarify these residual aspects at a later time.

## CREDITS

The author wishes to thank Luca Lussari, Department of Mathematics and Physics of Università Sacro Cuore di Brescia - Italy, for his suggestions and observations.

## BIBLIOGRAPHIC REFERENCES